\documentstyle[12pt,epsfig]{article}
\newcommand{\beeq}{\begin{equation}}
\newcommand{\eneq}{\end{equation}}
\newcommand{\beeqar}{\begin{eqnarray}}
\newcommand{\eneqar}{\end{eqnarray}}

\begin{document}
\begin{center}
{\bf\large 
Regularisation of Chiral Gauge Theories}\\
By\\
N.D. Hari Dass\\
Institute of Mathematical Sciences, C.I.T Campus, Chennai 600 113\\
\vskip 0.5in
{\bf \large Abstract}\\
\end{center}
This article gives a review of the topic of regularising chiral gauge theories and is aimed at a
general audience.It begins by clarifying the meaning of chirality and goes on to discussing chiral
projections in field theory, parity violation and the distinction between vector and chiral field
theories.It then discusses the standard model of electroweak interactions from the perspective of
chirality. It also reviews at length the phenomenon of anomalies in quantum field theories including
the intuitive understanding of anomalies based on the Dirac sea picture as given by Nielsen and
Ninomiya.It then raises the issue of a non-perturbative and constructive definition of the standard model
as well as the importance of such formulations. The second Nielsen-Ninomiya theorem about the impossibility
of
regularising chiral gauge theories under some general assumptions is also discussed. After a brief review of lattice regularisation of field
theories, it discusses the issue of fermions on the lattice with special emphasis on the problem of
species doubling. The implications of these problems to introducing chiral fermions on the lattice
as well as the interpretations of anomalies within the lattice formulations and the lattice Dirac sea
picture are then discussed.Finally the difficulties of formulating the standard model on the lattice are illustrated
through detailed discussions of the Wilson-Yukawa method, the domain wall fermions method and the recently
popular Ginsparg-Wilson method.
\section{Introduction}
`Chirality' simply means handedness. Handedness in its most basic meaning means
a correlation between circular motion and linear motion. For example,when the
head of a corkscrew is given a rotation, the tip of the screw moves forward or
backward depending on which way the head is rotated. A quantity representing
this correlation is the scalar product of angular momentum $\vec J$,the generator
of rotations and momentum $\vec P$, the generator of translations. Actually the
magnitudes of the angular momentum and linear momentum are irrelevant for
quantifying the desired correlation. This
quantity $\vec J\cdot\vec P$ is called `helicity' of the particle.

For massive particles, it can easily be seen that helicity depends on the
initial frame of reference. If the inertial frame is the rest frame of the
massive particle, helicity is not even defined. The direction of the linear
momentum can change depending on the frame while the direction of angular
momentum does not. Hence helicity can take both signs.

But for massless particles helicity has a Lorentz-invariant meaning. No rest 
frame is available to these particles and consequently the direction of the momentum
can not be reversed by a change of frames. This fact is of special significance
in the case of fermions. Specifically, in $d=4$,the Dirac-Weyl equation
describing massless fermions is invariant under the `$\gamma_5$-transformation:
\beeq
\delta\psi = i\epsilon\gamma_5\psi
\eneq
As a consequence,$\gamma_5$ is conserved and its eigenvalues are the particle
helicities. From now onwards we shall call the $\gamma_5$-eigenvalues
`chirality'.

It also follows that we can introduce the chirality projection operators
$P_{\pm}={1\over 2}(1\pm\gamma_5)$ and write
\beeqar
\psi_L & = & P_+\psi \nonumber\\
\psi_R & = & P_-\psi 
\eneqar
\section{Parity Violation}
A development of great significance in physics was the discovery of parity violation.
It was found that in $\beta$-decay,the electron was emitted in a direction
predominantly anti-parallel to the direction of nuclear spin.In a landmark
development Sudarshan,Marshak,Feynman,Gell-mann and Sakurai showed that the
weak interaction currents were of the $V-A$ type. We will see that this has
had profound implications for the conceptual developments in particle physics.
In its impact and timelessness this discovery should rank along with Galileo's
inertia, Einstein's equivalence principle etc.

Mathematically speaking the weak interaction current has the form
\beeq 
J_{\mu}^{+} = \bar p_L \gamma_{\mu} n_L
\eneq
The remarkable feature of this current is that it is made of only the 
lefthanded fields. In fact it is a property of $\gamma_{\mu}$-interactions
that they preserve chirality.It is instructive to compare the structure of 
the electromagnetic current also expressed in terms of the chiral components:
\beeq 
J_{\mu}^{el} = 
\bar \psi_L \gamma_{\mu} \psi_L+
\bar \psi_R \gamma_{\mu} \psi_R
\eneq
Again the current does not mix the L and R components.However,in the electromagnetic
current the L and R fields occur on an equal footing, reflecting the parity
conserving nature of the electromagnetic interactions.Such theories will
henceforth be referred to as {\bf Vector} theories in contrast to the theory of
weak interactions which shall be called {\bf Chiral}.

As far as the structure of these currents are concerned, it appears possible
to treat the L and R fields as independent species of particles. But as mentioned
before, chirality does not have a Lorentz-invariant meaning for massive
particles. In fact the mass term in a Lagrangean expressed interms of L and R
fields looks like
\beeq
m\bar\psi \psi = m(\bar\psi_L \psi_R + \bar \psi_R \psi_L)
\eneq
Thus inertia can be viewed as merely an interaction that switches L and R
species! This is indeed a major paradigm shift in elementary particle theory.
\section{The Standard Model of Weak and Electromagnetic Interactions}
The next major development of relevance is the standard model which unified
in one step the old Fermi theory of weak interactions, the $V-A$ structure of
weak currents inspired by the observed parity violation in beta decay,and
remarkably,the electromagnetic interactions. Other highlights of this theory
were the 'dynamical' origin of masses, existence of weak neutral currents
and from a theoretical point of view, renormalisability.

The construction of the standard model is based on the symmetry group $SU(2)_L
\times
U(1)$.Restricting ourselves to the leptonic sector(electron,neutrino) for
convenience, the basic fields are taken to be $(e_L,{\nu_e}_L)$ transforming
as a doublet under $SU(2)_L$,and $e_R$ transforming as a singlet under $SU(2)_L$.
Both the left-handed doublet and the righthanded singlet transform non-trivially
under $U(1)$.Furthermore, the $U(1)$ charges of the L and R fields are
different.The vector currents $\bar\psi_L\gamma_{\mu}\vec\tau\psi_L$ transforming
as a triplet under $SU(2)_L$ and $\bar\psi_L\gamma_{\mu}\psi_L$,$\bar\psi_R
\gamma_{\mu}\psi_R$ couple to the gauge fields $\vec W_{\mu}$ and $B_{\mu}$.
The neutral currents couple to the $U(1)$-gauge field in proportion to the 
hypercharges of the L and R fields.

It should be emphasised here that the parity violations are "put in by hand"
in the standard model.This is realised by ascribing very different properties
(like hypercharges,$SU(2)_L$ representations) to the R and L fields.These
difference do not arise as manifestations of any deeper dynamics.Whenever
l and R fields transform differently under the group, one says that there
are complex representations of the group.

Associated with the gauge fields are the following {\bf local} gauge transformations
under which the theory is invariant. In addition to the fermionic fields and
gauge fields, the standard model also has the so called Higgs field which
transforms as a complex $SU(2)_L$ doublet.The four real components of this
complex doublet can also be arranged as a $2\times 2$ matrix $\Phi$:
\beeqar
\phi^0+\phi^3 &~~~~~& \phi^1+i\phi^2 \nonumber\\
\phi^1-i\phi^2 &~~~~~& \phi^0+i\phi^3 
\eneqar
Likewise,the gauge fields $\vec W_{\mu}$ can be equivalently represented as
a matrix $\bf W_{\mu} = \vec W_{\mu}\cdot\vec \tau$. In terms of these fields
and the fermionic fields the gauge transformations are:
\beeqar
B^{\prime}_{\mu}(x) &=& B_{\mu}(x)+\partial_{\mu}\theta(x)\nonumber\\
\vec {\bf W}^{\prime}_{\mu}(x) &=& g(x){\bf W_{\mu}} g^{-1}(x)
-i\partial_{\mu}g(x)\cdot g^{-1}(x) \nonumber\\ 
\psi_L^{\prime}(x)&=& g(x)e^{iy_L\theta(x)}\psi_L(x)\nonumber\\
\psi_R^{\prime}(x)&=& e^{iy_R\theta(x)}\psi_L(x)\nonumber\\
\Phi^{\prime}(x)&=& e^{i(y_L-y_R)\theta(x)}g(x)\Phi
\eneqar
These transformation rules allow for gauge-invariant interactions
of the type
\beeq
{\cal L}_{Yukawa} = g_Y \bar\psi_L\Phi\psi_R+h.c
\eneq
It is easily recognised that the same interaction can generate masses
for the fermions(see eq()) if the Higgs field $\Phi$ develops a vacuum
expectation value.But that would mean that the {\bf global} part of 
$SU(2)_L\times U(1)$ would be spontaneously broken and by the Goldstone theorem
there ought to be massless Goldstone bosons equalling {\bf at least}
the number of broken generators.The latter is estimated on noting that
$SU(2)_L$-rotations about the direction along which the VEV of $\Phi$
points still leave the action invariant.Thus there are three broken generators
and in perturbative analysis one finds three Goldstone bosons. Because 
the global invariance is elevated to a local one, the relevant phenomenon
is the Anderson-Higgs mechanism by which three of the vector bosons
acquire masses and the gauge boson corresponding to the unbroken generator
is identified as the massless photon.This is inded the conceptual economy
of the standard model which , while unifying weak and electromagnetic interactions,
naturally realises their importance,namely,the difference in the ranges of the
interactions.

As already stressed, the standard model is perturbatively {\bf renormalisable}. This hinges
on the delicate fact that breaking the symmetries spontaneously does not
spoil renormalisability and that the symmetric version of the theory is
renormalisable.In fact one of the biggest stumbling blocks towards the
construction of a field theory of weak interactions was the non-renormalisability of 
{\bf generic} massive Yang-mills theories.Establishing this consists of first
regularising the theory , then checking all the Ward identities for 
the regularised theory and finally using definitions of an optimal set of
observables to "renormalise". One has the choice of either regularising
the theory maintaining all the symmetries(if possible),in which case the
Ward identities are automatically satisfied in the regularised theory, or,
using non-invariant regularisation schemes and adding the requisite 
(non-invariant)counter terms to realise the Ward identities. 

One of the popular techniques for regularisation is the so-called dimensional
regularisation where the dimension of space-time is taken to be $n=4-\epsilon$
and removing cut-off is equivalent to taking the limit $\epsilon\rightarrow 0$.
This way of regularising has the advantage that it is manifestly gauge invariant.

But already at this stage chirality begins to pose some problems.The origin
of this difficulty lies in the fact that chirality is a very dimension
dependent concept i.e in odd dimensions there are no Weyl fermions.More
explicitly,$\gamma_5$ which enters the chiral projection operators(in $d=4$)
is given by $\gamma_5=\gamma_0\gamma_1\gamma_2\gamma_3$ and it is clear that
there is no straightforward way of generalising this to arbitrary dimensions.
However, there is a procedure due to Breitenlohner and Maison,which seems to
succesfully address this question ,at least in low orders of perturbation theory.

\subsection{A Caveat : The Anomalies}
The renormalisability of the standard model hinges on a caveat of the absence
are cancellation of the so called 'anomalies'.Conceptually,anomalies will play
a vital role in the rest of our discussions, so it is worth our while to
examine them carefully.Anomaly is an effect in Quantum Field Theory whereby
a symmetry is destroyed by the quantum fluctuations.It is highly counter-
intuitive as it is not clear why quantum fluctuations should invalidate
conservation laws.

On closer examination,anomalies can be understood as arising because of
infinite number of degrees of freedom characterstic of all field theories.It
should be recalled that the occurrence of divergences in QFT necessitating
the renormalisation procedure is also a consequence of these infinitely many
degrees of freedom.In the case of a theory like Quantum Electrodynamics or
Quantum Yang-Mills theory, regularisation and renormalisation can be carried
out preserving the corresponding gauge invariances.If, however, we consider
a theory with several invariances at the so-called 'classical' level,the
likelihood of there being no way of regularising the theory maintaining all
the ward identities can not be ruled out.Then a choice has to be made to
give up some of the invariances. This in a nutshell is the basis of the
anomalies.

Before we analyse the anomalies in depth,it is relevant to point out that not 
all invariances need be symmetries in the Wignerian sense,which,among other
things,would associate degeneracies with symmetries.Therefore,contrary to the
oft-used language,local gauge invariances are not symmetries and hence the
expression 'gauge symmetry' is an abuse of language.What local gauge invariances
represent are statements about the number of degrees freedom or more
precisely,they specify the physical configuration space of the theory.A rigid
or global gauge invariance,on the other hand, is a symmetry of the theory.
Instead of leading to degeneracies,it leads to superselection rules.

Let us now take a more detailed look at the anomalies.For that purpose,let
us consider a theory which is classically(in the sense of ignoring the
effect of quantum fluctuations) is invariant under
\beeqar
\delta\psi(x) &=& \alpha(x)\psi(x)\nonumber\\
\delta A_{\mu} &=&\partial_{\mu}\alpha(x)\nonumber\\
\delta\psi(x)&=&\beta\psi(x)
\eneqar
with the associated conservation laws
\beeqar
\partial_{\mu}j^{\mu}&=&0\nonumber\\
\partial_{\mu}j^5_{\mu}&=&0
\eneqar

The big surprise was that no regularisation scheme could be found that
maintained the ward identities corresponding to both the gauge transformations
($\alpha (x)$ - transformations) and the chiral transformations($\beta$-transformations).
Consequently if $\partial_{\mu} j^{\mu}_{ren}=0$,then $\partial_{\mu}j^{\mu}
_{5,ren}\ne 0$.

In the example considered above,chiral symmetry was global. In the standard
model,the invariances are local.Also,for simplicity we only considered Abelian
transformations.But in the standard model we have both Abelian and non-Abelian
transformations.It turns out that when the non-Abelian group is $SU(2)$,
there can not be any non-conservation of the non-Abelian currents(in $d=4$)
and one should only worry about the $U(1)$ anomaly.

It is clear that when the current coupling to a gauge field is anomalous,
gauge invariance is lost.As stated earlier,gauge invariance can be viewed
as a statement about the degrees of freedom of a theory.Thus if the object
is to construct a consistent theory with the required number of degrees 
of freedom,anomalies would render such a theory sick.

Indeed the general folklore was that anomalous gauge theories are sick and
ill-defined.If ,however,we take the view point that the true degrees of freedom
could be larger than the naive count of degrees of freedom,anomalous gauge
theories could in principle be consistent.

That this could indeed be the case was shown by Jackiw and Rajaraman.They
showed that in $d=2$ anomalous gauge theories can indeed be consistent,and
the price for the consistency were additional degrees of freedom.Whether or
not the same thing works in higher dimensions is still an open issue,though
the chances seem remote. 

Now the caveat in proofs of renormalisability is that the presence of
anomalies leads to additional sources of divergences which can not be absorbed
into the allowed set of counterterms.This was first pointed out by Gross and Jackiw.

The miracle of the standard model is that the full theory including quarks and leptons
is actually anomaly free and the above-mentioned caveat is no longer of any
concern.
\subsection{A Non-perturbative Definition of the Standard Model?}
So far our analysis of the standard model, a chiral gauge theory, has been
perturbative. But it is well known that perturbative analyses can be highly
misleading as for example in the $\lambda \phi^4$ theory in $d=4$ where perturbation
theory yields a non-trivial S-matrix but a fully non-perturbative analysis
shows the theory to be trivial(technically the proofs are still a bit
incomplete in $d=4$).Of course there are indications of this theory being
problematic through the appearance of so-called Landau ghosts in
perturbative analalysis. But it is not clear whether these are artefacts of perturbation
theory.

In the case of Quantum Electrodynamics,predictions of perturbation theory
are very well borne out experimentally.That theory too suffers from the
presence of Landau ghosts, so one can not immediately conclude from the presence of
Landau singularities that perturbative analysis is unreliable.

In the case of the standard model also, predictions of perturbation theory
are in excellent agreement with observations.So perturbative analysis is
perhaps not as misleading as in the case of $d=4 \lambda\phi^4$ theory.

Nevertheless a non-perturbative formulation of the theory that is mathematically
well defined is always desirable.From the point of view of confronting the
theory with experiments also such a formulation is desirable as it will make
hitherto inaccessible aspects of the theory amenable for verification.From 
a matter of principle also one should insist on such a formulation because
for the theory to make sense it should be well defined in all regions of its
parameter space i.e both perturbative and non-perturbative regions.

So we pose the following two questions about the standard model:\\

{\bf 1. Is there a nonperturbative definition of the theory?

2. If so, can that definition be a constructive one?}\\

Before answering these questions it is instructive to take another look
at the anomalies. We shall closely follow the reasonings of Nielsen and
Ninomiya\cite{nn1},who wanted to arrive at an 'intuitive understanding' of the
anomalies.In particular, to clarify the mystery of "how a classical
conservation law disappears quantum mechanically?".

Following them let us start with Weyl-particles in $1+1$ dimensions.The
dispersion relations are given by
\beeq
\omega = \pm p
\eneq
where the $\pm$ refere respectively to right and left movers.Let these
Weyl particles carry charges and consider the action of an electric field
$F_{01}$.Without loss of generality let $qF_{01} > 0$ where q is the charge
of the Weyl particles.
\subsection{Single Particle Picture}
If one restricted attention to the single particle sector with positive energy,
t is easy to see that under the influence of the electric field, the particle
momentum will steadily increase with time. For the right-movers this will imply
a steadily increasing energy while for the left-movers the energy steadily decreases.
The motion is along the dispersion curve, as shown in the figures below.This
is also called a "spectral flow".\\
\begin{figure}[htb]
\begin{center}
\mbox{\epsfig{file=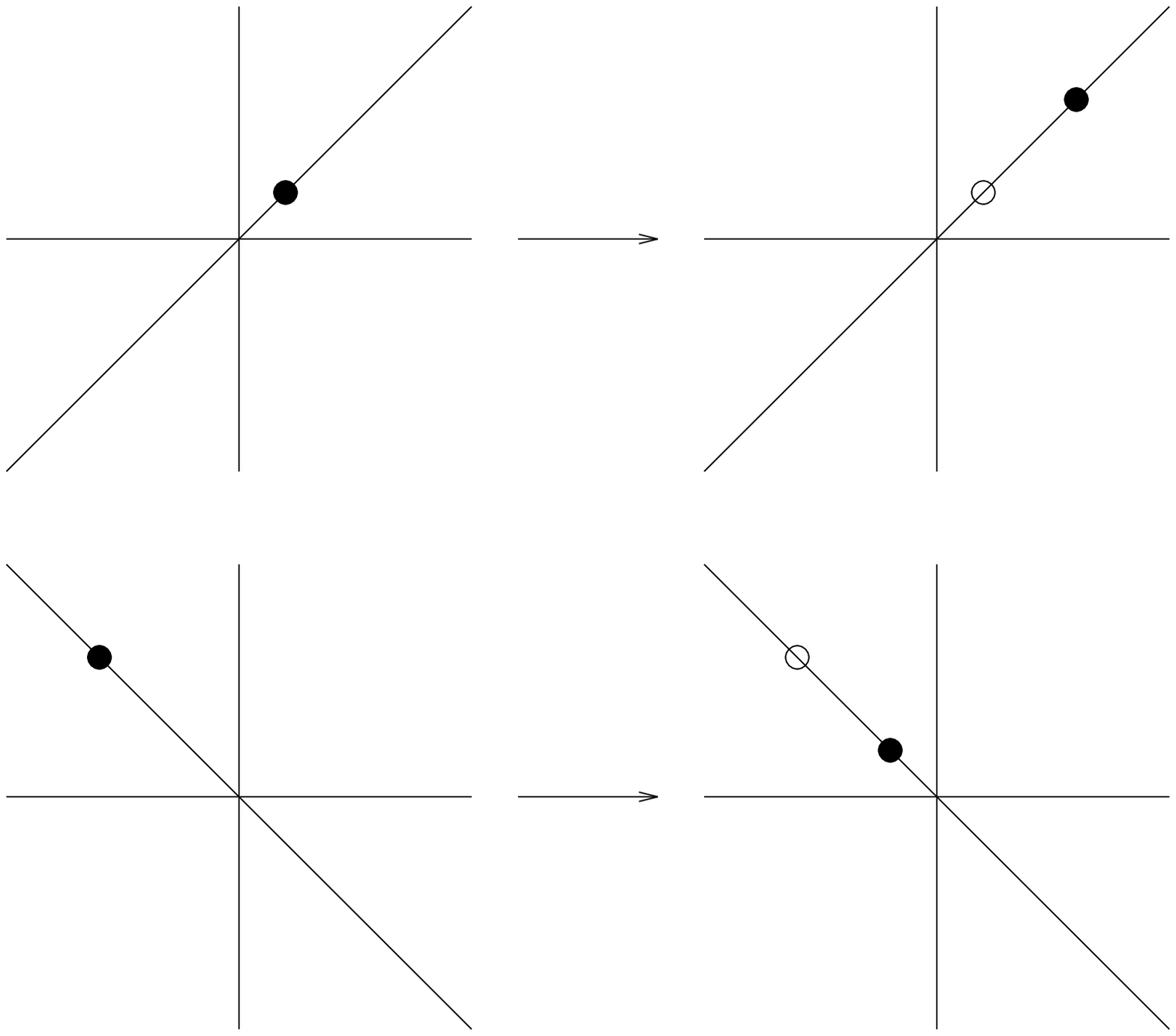,width=3truecm,height=2truecm,angle=-90}}
\caption{Single Particle Picture
}
\label{Fig 1.}
\end{center}
\end{figure}
Clearly no net chirality is produced;by net
chirality we mean the difference in the number of right-movers and left-movers.
It is also obvious that this picture remains if we consider a collection of
left and right-movers where each individual particle is of positive energy.This
is essentially the "classical" picture.

It is also clear that the above picture holds irrespective of whether the dynamics
conserves parity or not as long as it conseves chirality.More precisely,instead
of the parity conserving electromagnetic interaction
\beeq
{\cal L}_{em} = 
(\bar\psi_L\gamma_{\mu}\psi_L
+\bar\psi_R\gamma_{\mu}\psi_R
A_{\mu}
\eneq
one had considered the parity violating interaction
\beeq
{\cal L} = 
(g_1\bar\psi_L\gamma_{\mu}\psi_L
+g_2\bar\psi_R\gamma_{\mu}\psi_R
B_{\mu}
\eneq
net chirality would still be preserved even though the rates at which the left and right-movers
move along the spectral curve would be different.If on the other hand,the dynamics was itself
chirality non-conserving even if parity-conserving as in
\beeq
{\cal L} = \bar\psi_L\Phi\psi_R + h.c
\eneq
there would be net chirality production in proportion to the original chirality.
\subsection{Influence of The Dirac Sea}
However, we know that the single particle picture is quantum mechanically
incomplete both for fermions and bosons. In the former case,one possible
resolution is to invoke the concept of the "Dirac Sea" whereby all the
physically undesirable negative energy states are completely filled. Pauli
exclusion principle would then forbid transitions from the positive energy
states into negative energy states thereby stabilising the positive energy
states.However, it would always be possible to lift a particle in the Dirac
sea to a positive energy state leaving behind a "hole" in the Dirac sea
which would have positive energy relative to the Dirac sea state(vacuum state)
and be oppositely charged compared to the electron.This is the "positron" state,
popularly called the anti-particle state.

Though modern formulations of field theory exist which do not explicitly invoke
the concept of the Dirac Sea,it is nevertheless instructive to analyse field
theoretic phenomena in terms of the visually more transparent Dirac Sea.

Following Nielsen and Ninomiya, let us now take a fresh look at the effect
of the electromagnetic fields on charged Weyl particles.The vacuum or the
Dirac Sea has all the Right-handed particle states with negative momentum
and all the Left-handed particle states with positive momentum fully occupied,
as shown in the figure below:
\begin{figure}[htb]
\begin{center}
\mbox{\epsfig{file=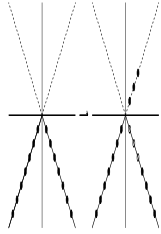,width=3truecm,height=2truecm,angle=-90}}
\caption{The Dirac Sea Picture
}
\label{Fig 2.}
\end{center}
\end{figure}
Using the spectral flow picture it is easy to see that when the electric field
is applied on the vacuum state,there will be a steady creation of R-particles
and at the same time a steady depetion of L-particles. Thus there will be net creation of
chirality and the axial current $j_{\mu}^5$ is no longer conserved! But since the
number of "holes" created is the same as the number of "particles" created,
total electric charge is indeed conserved.

Thus we get a qualitative understanding of axial anomalies through the Dirac
Sea picture.Actually even quantitatively the correct anomaly follows.If the
electric field has constant value E over a region of length $l$,one has
\beeqar
{dn_L\over dt} & = & ~l{qE\over h}\nonumber\\
{dn_R\over dt} & = & -l{qE\over h}
\eneqar
where $h$ is the Planck's constant.These equations immediately imply
\beeq
\partial_{\mu} j_{\mu}^5~=~{q\over \pi}E
\eneq
which is the correct form of the anomaly.
\subsection{The Four Dimensional Case}
The extension of the above mentioned arguments to higher even-dimensional
cases is straight forward.Let us consider the $d=4$ case as an example.The
Weyl equation now reads:
\beeq
i\gamma^{\mu}D_{\mu}\psi~=~0~~~~~~~~~~D_{\mu}~=~\partial_{\mu}-iqA_{\mu}
\eneq
Iterating this equation twice one gets
\beeq
\{-D_{\mu}D^{\mu}+{i\over 2}[\gamma^{\mu},\gamma^{\nu}]F_{\mu\nu}\}\psi=0
\eneq
Let us consider those field configurations for which $F_{\mu\nu}F^{\mu\nu}\ne 0$.
By choosing an appropriate Lorentz-frame,it is possible to make $F_{01}\ne 0$ and 
$F_{23}\ne 0$ with all other components of $F_{\mu\nu}$ vanishing.On noting
that$[\gamma_0,\gamma_1]$, $[\gamma_2,\gamma_3]$ and $\gamma_5=i\gamma_0\gamma_1
\gamma_2\gamma_3$ form a mutually commuting set,the spectrum can be labelled
by their simultaneous eigenvalues.In the magnetic field $F_{23}$,the charged
particles form Landau levels with degeneracy $\pi {2F_{23}(Area)_{23}\over (2\pi)^2}$.
Now because of the special choice of directions of the electric and magnetic
fields,the $3+1$-dimensional problem can be treated as if it were a $1+1$-dimensional
problem with the additional degeneracy.Thus
\beeq
{d(n_R-n_l)\over dt} = {Vol\over(2\pi)^2}qF_{01}F_{23}
\eneq
leading to
\beeq
\partial_{\mu}j_{\mu}^5={q\over (2\pi)^2}\epsilon_{\mu\nu\rho\sigma}F^{\mu\nu}F^{\rho\sigma}
\eneq
which is again the correct formula for the $U(1)$-anomaly.

It would be interesting to work out the non-abelian versions of these results.

{\bf Lesson}: Anomaly is the continuing pumping out of the (infinite) Dirac
Sea!Stated differently,it is the bottomlessness of the Dirac sea that allows
pumping of net chirality without any paying any price.
\subsection{Regularising the Dirac Sea}
An infinitely deep Dirac Sea is clearly an unphysical idea arising out of the
idealisation that energy and momenta can take arbitrary values.Another way of stating the
crux of this matter is that the infinite Dirac Sea is tantamount to infinitely
many degrees of freedom. It should be recalled that the ultraviolet divergences
arising in Quantum Field Theory, necessitating the renormalisation procedure,
always have their source in the assumption of infinitely many degrees of freedom.In
fact,mathematically the theory is meaningless under these circumstances.Therefore
one first works with a regularised version of the theory which is mathematically
well defined. A regularised version of a quantum field theory is a suitably and
consistently truncated version of that theory.

Clearly,the Dirac Sea has also to be regularised or in other words provided with a 
"bottom".Alternatively, regularisation in QFT throws away the processes happening
at the bottom of the sea as events occurring at a very high energy scale and
not relevant to the physics at low energy scales.In particular, the contribution to the net chirality by the inflow
at the bottom of the sea is ignored in the process of regularisation.
\begin{figure}[htb]
\begin{center}
\mbox{\epsfig{file=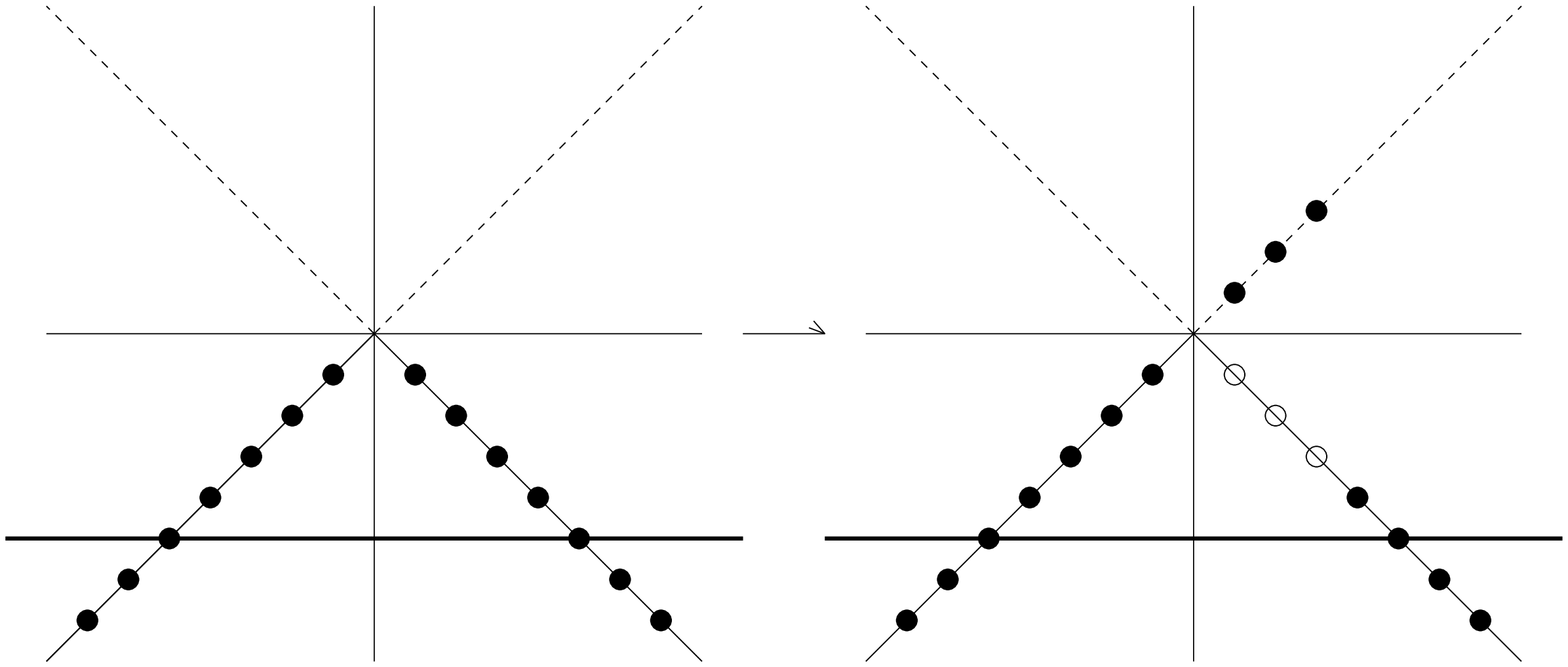,width=3truecm,height=2truecm,angle=-90}}
\caption{Truncated Dirac Sea
}
\label{Fig 3.}
\end{center}
\end{figure}
Since the anomaly was seen to be heavily dependent on the Dirac sea being infinite,one
may wonder as to the fate of the anomalies if one were to regulate the 
Dirac Sea.

Indeed it is clear that when one regularises the Dirac sea,one {\it unavoidably}
introduces dynamics which breaks the conservation of chirality {\bf unless one invents a very special
type of regularisation}.We shall see later that Lattice regularisation is
special in this sense.Of course, care should be exercised as to how exactly
the Dirac Sea is regularised as otherwise even electric charge and perhaps
momentum also may not be conserved.
\subsection{What Constitutes a Regularisation?}
Essentially a regularisation is replacing the original continuum (but
mathematically ill-defined) theory by one that is a very good approximation
to it at large distance scales but is finite(and hence mathematically 
well-defined).

It is a very reasonable premise that the truly fundamental theory is mathematically
well-defined and therefore {\bf finite}.From a physical stand point also,it is
reasonable to expect it to be finite as infinity is an idealisation never to
be realised under actual physical circumstances. It is important to emphasise that by
this reckoning the truly fundamental theory should be finite and not just
{\bf renormalisable} as in the latter case one hides some ignorance through the
renormalisation procedure and can not qualify to be a truly fundamental theory.

{\bf Thus the finite fundamental theory can be thought of as a regularisation
for its lower energy effective theory.}

Then the paradox of regularisation,namely,the unavoidable breaking of
chirality conservation in a regularised theory(generic),will also be a paradox
for any truly fundamental theory.

\section{The Second Nielsen-Ninomiya Theorem}\cite{nn2}
Nielsen and Ninomiya proved two important theorems in the context of chirality.
In the first they showed that it was impossible to put neutrinoes on a lattice
without explicitly breaking chiral invariance. In the equally fundamental
second theorem they showed that under rather general conditions it would be
impossible to regularise {\bf Chiral Gauge Theories}.In our discussion till now
the gauge aspects have only been implicit to the extent that a classically conserved chiral
charge was assumed but its coupling to a gauge field was not considered.The  
second NN-theorem deals with the difficulties of regularisation of chiral
gauge theories.

{\bf Statement}: It is not possible to simultaneously fulfill all of the
following:

{\bf 1. Fundamental Regularisation}

The fundamental theory is taken to be a finite theory.

{\bf 2. Parity Violation}

Different number of right and left-handed species for given representations
of symmetry groups.

{\bf 3. Exact Gauge Invariance}

This eans gauge invariance should be valid also at the regularisation scale.
Veltman has argued that even "small" violations of gauge invariance at short distances can have large 
effects on S-matrix elements. This is also in conformity with our earlier
remarks on gauge invariance being a specification of the degrees of freedom
and that there can not be any meaning to the breaking of this invariance.

{\bf 4. Bilinearity}
A technical assumption is made about the action being bilinear in the Weyl-fields.

\subsection{Are Superstrings an Exception?}
As commented by Nielsen and Ninomiya, Superstring theories seem to offer a way
out of the no-go theorem. Superstring theories are finite, at least perturbatively. 
It is worth commenting on the sense in which these theories are finite.In
these theories only the spectrum and S-matrix are calculable. The S-matrix
is found to be finite in every order of perturbation theory without recourse to
any renormalisation.However,the spectrum contains an {\bf infinite} species
of particles and momenta can take arbitrarily large values. It would be interesting
to investigate whether non-perturbatively there are indeed {\bf finite} degrees of freedom
as befitting a truly fundamental theory.

Now the point is that some superstring theories have in their spectrum chiral
fermions in complex representations as well as gauge field coupled to them.

To this extent superstring theories seem to evade the no-go theorem.How exactly
do they achieve this?Is it that the no-go theorem is valid only in local field
theories?Though superstring theories are not local field theories, they do have
some locality properties in that the string-interactions are local. It will be
interesting to fully understand this issue. 

It should be emphasised that the possibility that some of these superstring
theories may not turn out to be phenomenologically succesful is of no consequence
to this discussion of matters of principle.

\section{Lattice Regularisation}
\begin{figure}[htb]
\begin{center}
\mbox{\epsfig{file=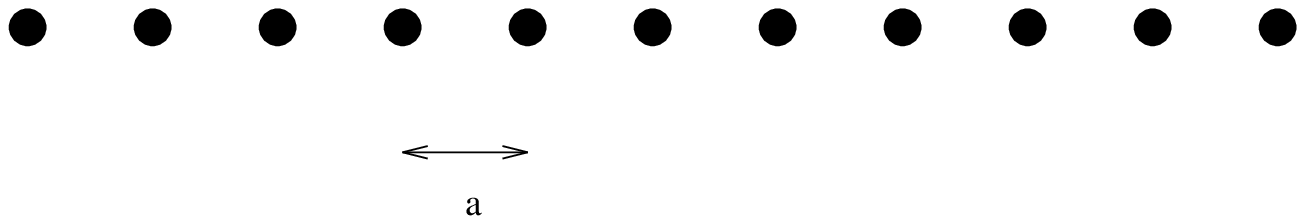,width=3truecm,height=2truecm,angle=-90}}
\caption{An One Dimensional Regular Lattice
}
\label{Fig 4.}
\end{center}
\end{figure}
In the lattice regularisation space-time is approximated by a discrete set of
points i.e $x^{\mu} \rightarrow n^{\mu}a$ where $n^{\mu}$ is an integer-valued four-vector
and $a$ is the lattice spacing.A scalar field $\phi(x)$,for example,is
represented by $\phi_{\bf n}$ where $\bf n$ stands for the four-vector.It turns
out to be useful to work with only dimensionless objects viz. $\phi_L=a\phi$ etc.
The derivatives of fields are replaced by finite differences. For example,
\beeq
\partial_{\mu} \phi(x) \rightarrow {\phi_{{\bf n}^\prime}-\phi_{\bf n}\over ma}
\eneq
where ${\bf n}^\prime-{\bf n} = m{\bf e}^{\mu}$ and ${\bf e}^{\mu}$ is the
unit vector in the $\mu$-direction.

It is quite clear that the choice of lattice i.e hypercubic,triangular etc
as well as the choice of finite difference chosen to approximate field derivatives
are arbitrary.It is believed that in the continuum limit these differences
should become irrelevant.

Some of the remarkable features of the lattice formulation are that it affords
a manifestly gauge-invariant regularisation even for non-Abelian gauge theories.
Furthermore,it produces a regularised theory while most regularisation schemes used 
in continuum quantum field theories regularise process by process as for example
in Pauli-Villars regularisation of Feynman diagrams.This also means that
lattice regularisation is a {\bf non-perturbative} regularisation.Therefore it
gives a non-perturbative formulation of the theory.

An important feature of Lattice regularisations is that momentum space is
compact and is topologically a d-torus if the quantum field theory is
formulated in d-spacetime dimensions. This will be seen to have a profound
impact on regularising chiral gauge theories.More precisely,the Euler characterstic
of the momentum space(Brilloin zone) is 0.
\subsection{Fermions on the Lattice}\cite{smit}
Consider the continuum Dirac equation
\beeq
i\gamma^{\mu}\partial_{\mu}\psi+m\psi = 0
\eneq
A possible candidate for the lattice equivalent of this is
\beeq
i\gamma^{\mu}\Delta_{\mu}\psi+m\psi = 0
\eneq
where the (forward)shift operator $\Delta_{\mu}$ is defined by
\beeq
\Delta_{\mu} f({\bf x}) = f({\bf x}+a{\bf e}_{\mu})-f({\bf x})
\eneq
In momentum space the continuum eqn(22) reads
\beeq
(i\gamma_{\mu}p_{\mu}+m)\psi=0
\eneq
while the lattice-Dirac eqn (23) reads
\beeq(i\gamma_{\mu}\sin p_{\mu} +m)\psi=0
\eneq
As was first pointed out by Smit and Wilson(independently),the difference 
between the two Dirac eqns is profound.To see this note that the $\sin$-function vanishes not only
at $p_{\mu}\simeq 0$ but also at $p_{\mu}\simeq \pi^A_{\mu}$ where $\pi^A_{\mu}$ are four momenta such that
the components are either $0$ or $\pi$(note that $p_{\mu},m$ in eqn(25,26) are
dimensionless).Indeed,$A$ takes values $1,..,16$ corresponding to four vectors
$(0,0,0,0)$,$(\pi,0,0,0)$(4 in number),$(\pi,\pi,0,0)$(6 in number),$(\pi,\pi,\pi,0)$
(4 in number) and $(\pi,\pi,\pi,\pi)$.If we expand $p_{\mu}$ around $\pi^A_{\mu}$ as
$p_{\mu}=\pi^A_{\mu}+q_{\mu}$ where $q_{\mu}$ are small,we would have
$\sin p_{\mu}= \pm \sin q_{\mu}$.It is quite easy to find nonsingular operators
$S_A$ such that
\beeq
S_A\gamma_{\mu}S_A^{-1}=\pm \gamma_{\mu}
\eneq
for every $\mu$ such that $\pi^A_{\mu}=\pi$. In fact for any $\rho$ such that
$\pi^A_{\rho}\ne\pi$,$S_A$ can be chosen to be $\gamma_{\rho}$.Now the lattice
Dirac eqn takes the form
\beeq
S_A(i\gamma_{\mu}\sin q_{\mu}+m)S_A^{-1}\psi =0
\eneq 
for every value of $A$.Alternatively,$\tilde\psi_{(A)}=S_A^{-1}\psi$ satisfy
\beeq
(i\gamma_{\mu}\sin q_{\mu}+m)\tilde\psi_{(A)} =0
\eneq 
Thus the lattice-Dirac eqn(23) actually represents 16 Dirac particles and in the
continuum limit,the lattice theory has the {\bf wrong} spectrum!
\section{Vector Gauge Theories on The Lattice}
Consider a Vector gauge theory like Quantum Chromodynamics(QCD) on the lattice.
This is vector like because the left and right-handed fields both transform
as the fundamental representation of $SU(3)$.The $SU(3)$ is also gauged, with
the interaction between the gauge fields $A^a_{\mu}$ and the quark fields given by
\beeq
{\cal L}_{fer,gauge}=A^a_{\mu}(
\bar\psi_L\gamma^{\mu}\tau^a\psi_L
+\bar\psi_R\gamma^{\mu}\tau^a\psi_R)
\eneq
Wilson proposed the following remedy for the problem of "species doubling"(16 Dirac
particles in place of 1,doubling in each space-time direction).He proposed
modifying the fermion lagrangean to
\beeq
{\cal L} = {1\over 2}\sum \bar\psi\gamma_{\mu}D_{\mu}\psi+m\sum \bar\psi\psi
-r\bar\psi D_{\mu}D^{\mu}\psi
\eneq
where
\beeq
D_{\mu}=U_{\mu}(x)\psi(x+{\bf e}_{\mu})-\psi(x)
\eneq
with $U_{\mu}(x)$ being the link variables(see McKellar's talk for details).
Wilson's modification has the effect of the replacement
\beeq
\gamma^{\mu}\sin p_{\mu}~~~\rightarrow~~~\gamma^{\mu}\sin p_{\mu}+r\sum_{\mu}
(1-\cos p_{\mu}
\eneq
in the lattice-Dirac eqn(23).For $p_{\mu}\simeq\pi^A_{\mu},A\ne 0$,the
added terms are $(2r,4r,6r,8r)$.Thus for $r\ne 0$ they have the effect of 
moving the masses of the doublers to $\simeq {1\over a}$ and hence infinity in the
continuum limit.Consequently the doublers can be made to decouple in the
continuum limit.

It is very important for this construction that the (added)Wilson term is
{\bf manifestly gauge-invariant}.
\subsection{Anomaly on The Lattice}
It is straight-forward to work out the divergence of the axial current in the
lattice regularisation.The result is
\beeq
\Delta_{\mu}j^{5}_{\mu}(x) = 2m\bar\psi(x)\gamma_5\psi(x)+{r\over a}(\bar\psi(x)
\gamma_5\psi(x)-{1\over 2}\sum_{\mu}\bar\psi(x)\gamma_5U_{\mu}\psi(x+{\bf e}_{\mu}))
\eneq
When $r=0$,i.e when the Wilson modification is not made,the anomaly is seen
to vanish exactly(we will explain the physical origin of this shortly) but
when $r\ne 0$ it can be shown that in the continuum limit $a\rightarrow 0$,the correct anomaly
is reproduced by the above equation.
\section{Chiral Fermions On The Lattice}
Species doubling,unlike in the case of vector gauge theories where it can be
handled quite satisfactorily,becomes really problematic in the case of chiral
gauge theories and it essentially makes it very difficult to lattice-regularise
such theories.However,very recently,a ray of hope appears to have emerged which will be
discussed in the last section.
Suppose we atart with
\beeq
{\cal L}_{chiral}={1\over 2}\sum \bar\psi_L\gamma_{\mu}D_{\mu}\psi_L
\eneq
where $\psi_L={1\over 2}(1+\gamma_5)\psi$ is the left-handed field in the 
continuum.
The naive expectation would be that the above lagrangean represents a single
species of a chiral(left-handed in this case) fermion.REcall that in the case of the Dirac fermion,the
naive expectation was belied by species doubling.What happens in the present case?Are there
also 16 chiral(left-handed) fermions?In fact what happens is far more subtle
and dangerous.

To see this recall that the transformation $S_A$ used to map the lattice-Dirac eqn to the same
form around all the $\pi^A_{\mu}$ changed the signs of those $\gamma_{\mu}$
such that $\mu$ were the directions where the components of $\pi^A$ were $\pi$.
This can be used to see the effect of $S_A$ on $\gamma_5=i\gamma_0\gamma_1\gamma_2
\gamma_3$.This is presented below
\beeqar
\Pi^0&=&(0,0,0,0)~~1~~~~~~~+,+,+,+~~~~\gamma_5\rightarrow\gamma_5\nonumber\\
\Pi^{\mu}&=&(\pi,0,0,0)~~4~~~~~~~-,+,+,+~~~~\gamma_5\rightarrow -\gamma_5\nonumber\\
\Pi^{\mu\nu}&=&(\pi,\pi,0,0)~~6~~~~~~~-,-,+,+~~~~\gamma_5\rightarrow\gamma_5\nonumber\\
\Pi^{\mu\nu\rho}&=&(\pi,\pi,\pi,0)~~4~~~~~~~-,-,-,+~~~~\gamma_5\rightarrow-\gamma_5\nonumber\\
\Pi^S&=&(\pi,\pi,\pi,\pi)~~1~~~~~~~-,-,-,-~~~~\gamma_5\rightarrow\gamma_5
\eneqar
Thus around the points $\pi^0,\pi{\mu\nu}$ and $\pi^S$ we do indeed have left-handed
modes but at $\pi^{\mu},\pi^{\mu\nu\rho}$ we actually have right-handed modes!
Taking into account the multiplicities of these points we see that eqn() naively thought
to represent {\bf one} left-handed field actually represents {\bf 16} fields
{\bf 8} of which are left-handed and {\bf 8} are right-handed!Not only have the
species been doubled,the naive {\bf parity asymmetric} situation actually
represents {\bf parity symmetric} situation! 

A more careful analysis reveals that the left and right-handed fields carry the
same representation of the gauge group. Further the doubling arranges for the
total axial charge to be zero i.e $\sum Q_5=0$.

Before analysing the implications of this striking result, let us explain the earlier mentioned result
in the context of vector theories,namely,the vanishing of the anomaly in the
$r=0$ case.The naive Dirac field can again be thought of as being composed of
a right-handed field and a left-handed field, both transforming identically
under the gauge group.As we have just seen,both the naive left and right-handed
fields are really 8 left and 8 right-handed fields on the lattice.The total
axial chrge being zero,there is no anomaly!
\subsection{Lattice Dirac Sea Picture}
At this stage it is instructive to understand the results of species doubling
and the vanishing of the anomaly from a Lattice Dirac Sea picture.This was done
by Ambjorn,Greensite and Peterson \cite{ldr} by extending the Dirac Sea ideas of Nielsen
\& Ninomiya, and Peskin.

As in the continuum case, it is useful to adopt the Hamiltonian version.The
lattice dispersion relation for Weyl particles reads
\beeq
\omega_k = \pm \sum_i \sin k_i
\eneq

This dispersion relation shows a dramatic change in the nature of the regularised
Dirac sea! A naive regularisation of the Dirac sea would have envisaged a sea with a bottom
but nevertheless such that the bottom is at a considerable depth. Consequently
one may have imagined that the happenings at the bottom of the sea are not of much relevance
to low energy phenomena.
\begin{figure}[htb]
\begin{center}
\mbox{\epsfig{file=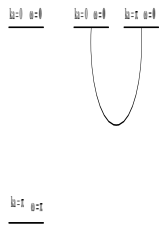,width=3truecm,height=2truecm,angle=-90}}
\caption{The Lattice Dirac Sea
}
\label{Fig 5.}
\end{center}
\end{figure}
However,the lattice dispersion relation () shows that the states with maximum
momentum in any of the directions(but zero momentum in the orthogonal directions)
are also at {\bf zero} energy and hence at the {\bf top} of the sea! What 
happens there is very much of consequence for low energy physics!

Other notable features of the lattice Dirac sea are (i) no gauge invariance violation
at the "bottom"of the sea and as discussed in the previous section,(ii) chirality is
flipped at half the number of "bottoms".Putting all these together one finds
that there is net pumping of chirality.
\subsection{Generic nature of Doubling}
One may wonder whether the species doubling that we have encountered is an
artefact of the way the fermions have been latticised and whether with some
luck one may find a way of latticising that would avoid species doubling.The
answer to this as given by the first nielsen-Ninomiya theorem is NO.According to
this theorem the occurrence of doubling is {\bf generic}.
\begin{figure}[htb]
\begin{center}
\mbox{\epsfig{file=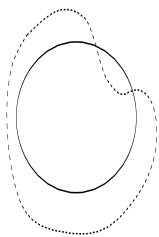,width=3truecm,height=2truecm,angle=-90}}
\caption{A Function With a Pair of Simple Zeroes on The Circle
}
\label{Fig 6.}
\end{center}
\end{figure}
The crux of this theorem is that the origin of species doubling is topological
in nature. As already stated before the momentum space(Brilloin Zone) is a d-torus
with Euler characterstic 0.The momentum space is also compact.This implies that
simple zeroes of a function i.e zeroes near which the function is linear,must
occur in pairs.It is the occurrence of such zeroes in pairs that translates into
species doubling.Thus chirality also must occur in pairs of opposite chirality.

Thus the problem of finding a discretisation that avoids species doubling
amounts to finding one whose associated Brilloin Zone has non-zero Euler Characterstic.
This appears to be a very difficult task.

\section{Standard Model on the Lattice?}
If the standard model could also be formulated on the lattice,we would have a non-perturbative
gauge-invariant formulation of it.But the lesson we have learnt is that species doubling
makes the theory vector-like without parity violation,unless a clever way is found to
avoid species doubling.This raises the following important question:

{\bf Can we move the doublers to the cut-off scale as was done 
for QCD by the Wilson Method?}
In the standard model the left-handed and the right-handed fields transform
differently under the gauge group.This precludes a bare mass-term for the
fermions.In fact,as discussed right in the beginning,masses for fermions 
are obtained through the Higg's mechanism.

For the same reasons,the Wilson mass term is also not gauge invariant in this
context.There have been many attempts to formulate chiral gauge theories on
tilhe lattice over the last 15 years.Most of them have failed in realising their
objectives.In the next section I'll describe three attempts at solving this problem.It is very
difficult to cover in a comprehensive way all the proposals that have
 been made to alleviate this problem.Many of the discussions in the literature are very technical.
Often,a conceptual separation is lacking of the problem of putting generic chiral gauge theories
on the lattice and the problem of putting the standard model on the lattice.
Due to a lack of space and time,
I have had to leave out the discussion of many interesting proposals
like the Rome Propsal\cite{roma}, the proposal of t'Hooft
\cite{hooft} to use different
regularisations for fermions and gauge fields,the proposal of Slavnov
\cite{slav},
the overlap
formalism of Narayanan and Neuberger \cite{over}
etc.(see \cite{shamir} for a more detailed coverage)
\section{Some Attempts To Put Standard-like Models On the Lattice}
\subsection{Wilson-Yukawa Models} \cite{hyuk}
Let us again recapitulate how the "mass terms" for fermions were generated in 
a gauge invariant manner within the standard model:the mass term
\beeq{\cal L}_{mass} = m(\bar\psi_L\psi_R+\bar\psi_R\psi_L)
\eneq
is not gauge invariant in a theory where the left and right-handed fields 
transform differently under the gauge group.The remedy was to consider the
gauge-invariant interaction term
\beeq
{\cal L}_{Yukawa} = g_Y\bar\psi_L\Phi\psi_R+h.c
\eneq
Gauge-invariance is achieved through a suitable transformation property of $\Phi$.
In the spontaneously broken phase of the theory,$\Phi$ develops a vacuum expectation value i.e
$<\Phi>=v$.Then the interaction term looks like
\beeq
{\cal L}_{Yukawa}=(g_Yv)\bar\psi_L\psi_R+h.c+....
\eneq
The fermion mass is given by $m_f=g_Yv$ and to get light fermions in the spectrum
one has to tune $g_Y$ appropriately.

One may attempt a similar trick to give masses of the order of cut-off to the
unwanted doublers.The idea is to generalise the Wilson mass term into a gauge
invariant Wilson-Yukawa term:
\beeq
{\cal L}_{Wilson-Yukawa}= \bar\psi_L\Phi(y-w\sum_{\mu}\partial_{\mu}\tilde
\partial_{\mu})\psi_R+h.c
\eneq
Here $w$ plays a role similar to $r$ in the Wilson term for vector theories;$\tilde
\partial_{\mu}$ is the backward shift operator.It should be remarked that this
construction is aimed more towards standard-model like theories rather 
than towards,say,chiral gauge theories on their own.

Now the doubler masses can be moved to infinity(cut-off scale) by taking the
limit $w\rightarrow \infty$. But this puts the theory in the strongly coupled
phase.A very careful study of this phase has been made by Golterman,Petcher,
Smit and others (for details, see \cite{hyuk}).Their conclusions are as follows:

i)Gauge singlets are formed as bound states of $\psi_L$ and $\Phi$.This in itself
need not be alarming in view of the so called t'Hooft Complimentarity picture
according to which there is no phase boundary between the Higgs and confining phases.
The spontaneously broken phase can be viewed in a manifestly gauge-invariant picture
and the massive gauge bosons would be viewed more like the gauge-invariant glue
ball states of pure QCD and would appear as bound states of the Higgs and gauge fields.
But in the context of chiral fermions the situation could be potentially problematic
as some authors claim a violation of complimentarity in this case.

ii)The real problem for the construction comes when one examines the interactions
in the theory.{\bf All interactions are seen to vanish in the $w\rightarrow\infty$}
limit.

iii)The doublers and the Right-handed particles decouple leaving behind a massive fermion
of mass $m_f=y$.Recall that $m_f=m^{phys}_fa$.Thus to get light fermions,$y$
has to be extemely fine-tuned.In fact $y$ has to vanish as $a$.The degree of
fine-tuning needed is much more severe than what is required to get light fermions
in the continuum standard model where $m_f=g_Yv$ because there $g_Y=m^{phys}_f/v^{phys}$.

With even very slight mismatch, {\bf all fermions decouple from the spectrum!}.
Even if fine-tuning could be achieved,the massless spectrum would consist of both
right-handed and left-handed particles.

Thus the Wilson-Yukawa approach does not work.
\subsection{Domain Wall Fermions}\cite{kap}
Another interesting proposal to put chiral fermions on the lattice was put forward by Kaplan
\cite{kap}.
His proposal consists in working on a five dimensional lattice to start with.
Since there are no chiral fermions in odd space-time dimensions and since there are
no problems in formulating vector theories on the lattice,this five dimensional
theory can be consistently and non-perturbatively formulated.
\begin{figure}[htb]
\begin{center}
\mbox{\epsfig{file=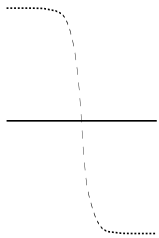,width=3truecm,height=2truecm,angle=-90}}
\caption{An Anti-Domain Wall }
\label{Fig 7.}
\end{center}
\end{figure}
Next he considers a four-dimensional domain wall to which chiral fermions are
constructed to be glued onto.The way this is accomplished is by considering
a 5-dimensional Dirac fermion whose mass depends on the 5th coordinate as follows:
\beeqar
m_5(x^5)&=&~~m~~~x^5>0\nonumber\\
        &=&~~0~~~x^5=0\nonumber\\
        &=&~-m~~~x^5<0
\eneqar
Here $m>0$ and the four-dimensional domain wall is at $x^5=0$.

The point is that normalisable solution of the Dirac equation is a {\bf single}
chiral fermion with $\gamma_5=1$ living on the domain wall.in the limit that 
the extent $L_5$ along the 5th direction tends to infinity. Narayanan and
Neuberger have shown that this picture can be given a purely four-dimensional
interpretation also.The anomaly in this picture arises as a Chern-Simons current flowing out
of the domain wall.
\subsubsection{Problems} 
With periodic boundary conditions (usually preferred in lattice studies) in
the 5th direction,one inevitably has an {\bf anti-domain wall} with a $\gamma_5=-1$
chiral fermion living on it.For finite values of $L_5$ there are contaminations
by unwanted chirality states.Coupling a gauge field to the domain wall chiral
fermion requires $d=4$ gauge fields only close to the domain wall and zero elsewhere.
Though there are some proposals on how to handle this,the fact that this implies gauge invariance violation
is not very encouraging.

It is of course possible to consider open boundary conditions in which case some of
these problems disappear.What results then is essentially the overlap formalism
of Narayanan and Neuberger.We shall not discuss that any further here as it is
rather technical and the program is still incomplete.
\subsection{Wilson-Ginsparg Method}
Ginsparg and Wilson (for details see \cite{nieder} ) gave a very interesting interpretation of what exact
chiral symmetry on the lattice means.They started with a cut-off theory with
exact chiral symmetry and consider block spinning transformations which are
chirally asymmetric.This way they obtain a coarse grained theory whose action is
chirally asymmetric but whose continuum theory is indeed chirally symmetric.In
this manner they found that the Grren's function of the coarse grained theory should
satisfy
\beeq
\gamma_5D+D\gamma_5=D\gamma_5D
\eneq
instead of the naive symmetric Green's function that would satisfy $\gamma_5D+D\gamma_5=0$.
Incorporating non-abelian gauge fields can be problematic.INterest in this idea was
revived by the observation of Hasenfratz that the fixed point action of QCD
satisfies the Ginsparg-Wilson relation.

Though the original work of Ginsparg and Wilson did not directly confront the
problem of regularising chiral gauge theories,much attention has been focussed in that
direction by an observation of Neuberger and Narayanan that the overlap formalism
produces a $D$ satisfying the Ginsparg-Wilson relation and by Lueschers
\cite{ginswils} claim of having constructed 
a $U(1)$ chiral gauge theory on the lattice.
On introducing the "lattice chiral transformations"
\beeq
\delta\psi=\gamma_5(1-{aD\over 2})\psi
\eneq
\beeq
\delta\bar\psi=\bar\psi(1-{aD\over 2})\gamma_5
\eneq
for abelian gauge theories and
\beeq
\delta\psi=T\gamma_5(1-{aD\over 2})\psi
\eneq
\beeq
\delta\bar\psi=\bar\psi(1-{aD\over 2})\gamma_5T
\eneq
for non-Abelian gauge transformations,it is easy to verify that
\beeq
\delta(\bar\psi D\psi)=0
\eneq
However,the measure for the functional integration over fermions is not invariant
under these transformations and the anomalies are reproduced this way very much the way the 
Fujikawa derivation of anomalies works in the continuum.

Though the Ginsparg-Wilson method offers at the moment the best hope for
putting chiral gauge theories on the lattice,there are still many open issues.
Even Lueschers\cite{ginswils} construction is a progress as far as matters of principle are concerned,but is not at
a stage where one can implement it.The non-Abelian extensions of it,an understanding of 
the GW construction in terms of the lattice Dirac sea,a better understanding of how the
Nielsen-Ninomiya theorem is circumvented are issues yet to be tackled.The
eventual goal would of course be an implementable lattice regularisation of the
standard model.

\end{document}